\documentclass[twoside,epsfig]{article}
\usepackage{fleqn,espcrc2,epsfig}

\usepackage{graphicx}
\usepackage[figuresright]{rotating}

\newcommand{\AmS}{{\protect\the\textfont2
  A\kern-.1667em\lower.5ex\hbox{M}\kern-.125emS}}

\title{Exact gauge invariant mass dependence of $\alpha_s$ through two loops}

\author{Michael Melles\address{Department of Physics,\\
        Durham University, \\ 
        Durham, DH1 3LE, U.K.}%
        \thanks{Research supported by the EU Fourth Framework Programme 
                `Training and Mobility of Researchers' through a Marie Curie
		Fellowship.}}
\begin{document}

\begin{abstract}
A physically defined QCD coupling parameter naturally incorporates massive quark
flavor thresholds in a gauge invariant, renormalization scale independent and
analytical way.
In this paper we summarize recent results for the finite-mass fermionic
corrections to the heavy quark potential through two loops leading to the
numerical solution of the physical and mass dependent 
Gell-Mann Low function. The decoupling-,
massless- and Abelian-limits are reproduced and an 
analytical fitting function is
obtained in the V-scheme. 
Thus the gauge invariant mass dependence of $\alpha_V$ is now known through
two loops. Possible applications in lattice analyses, heavy quark 
physics and effective charges are briefly discussed.
\vspace{1pc}
\end{abstract}

\maketitle

\section{Introduction}

Quark flavor thresholds in QCD are commonly treated within effective descriptions
in MS-like coupling definitions by imposing matching conditions at the quark thresholds \cite{bw,m}.
Thus quarks are considered infinitely heavy below and massless above $m_q$ and the coupling
is non-analytic at the thresholds. Real mass effects need to be calculated separately as higher
twist effects in the small and large mass limits. For the intermediate range an all orders
resummation of these expansions 
is necessary. In this paper we summarize recent results presented in Refs. \cite{me1,me2}
based on a physical coupling definition obtained from the static quark-antiquark potential \cite{s},
$V(Q^2,m^2)\equiv-4 \pi C_F \frac{\alpha_V(Q^2,m^2)}{Q^2}$, which naturally incorporates massive
quarks and where the scale $Q^2\equiv-q^2=\bf{q}^2$ is identified with exchanged momentum between the heavy
sources. A technical complication is that the massive Gell-Mann Low function can only be solved
numerically due to the complexity of the obtained results and that it is scheme dependent already
at one loop. The latter point can be ameliorated by expressing other physical charges through
$\alpha_V$ and using the conformal ansatz \cite{me2}.
We begin in the next section by reviewing the two-loop corrections including massive
quarks to $V(Q^2,m^2)$ and then discuss the solutions to the massive renormalization group
equations. Finally we briefly outline possible applications.
\begin{center}
\begin{figure}[p]
\epsfig{file=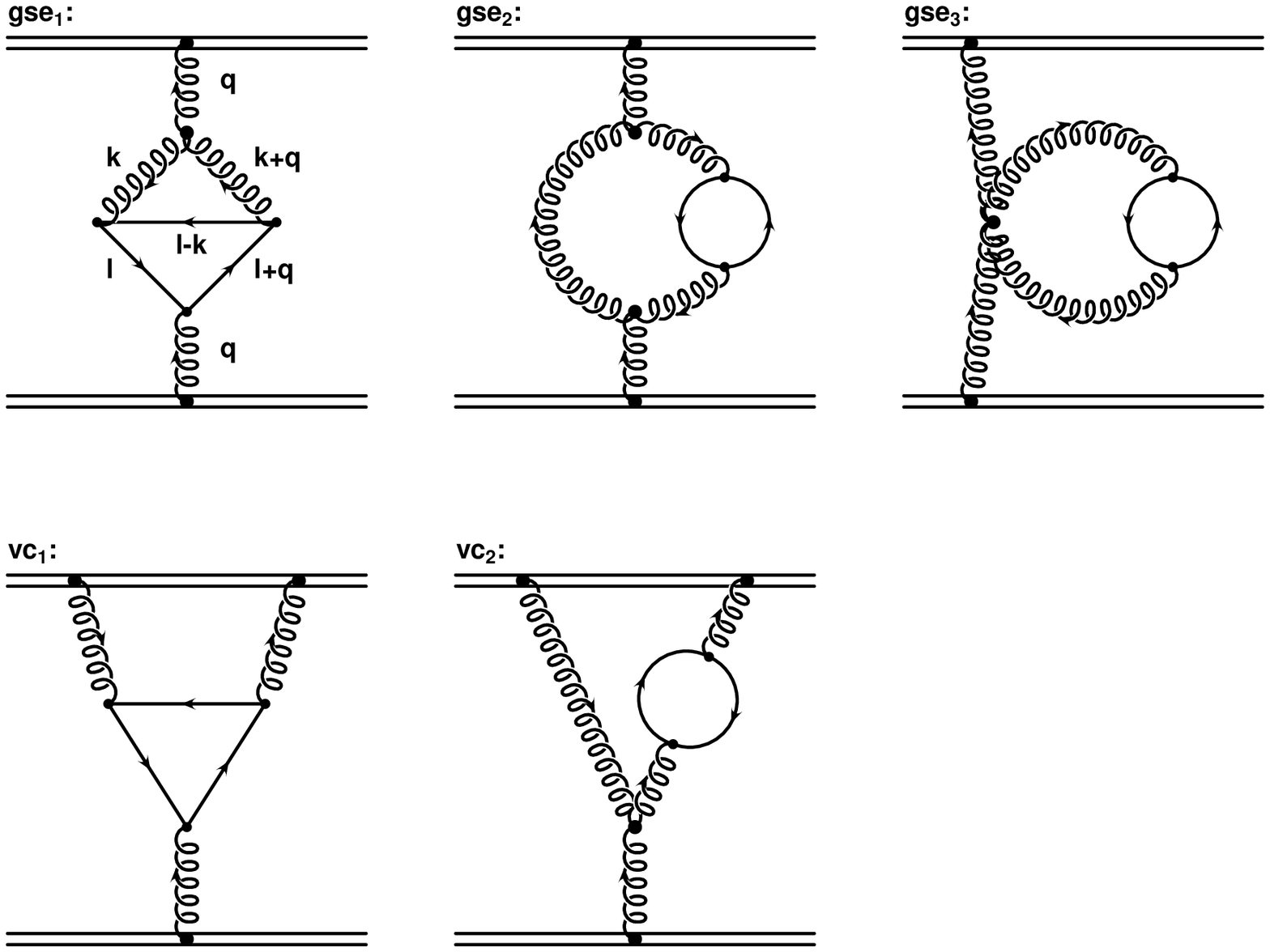,width=7.0cm}
\vspace{0.5cm} \\
\epsfig{file=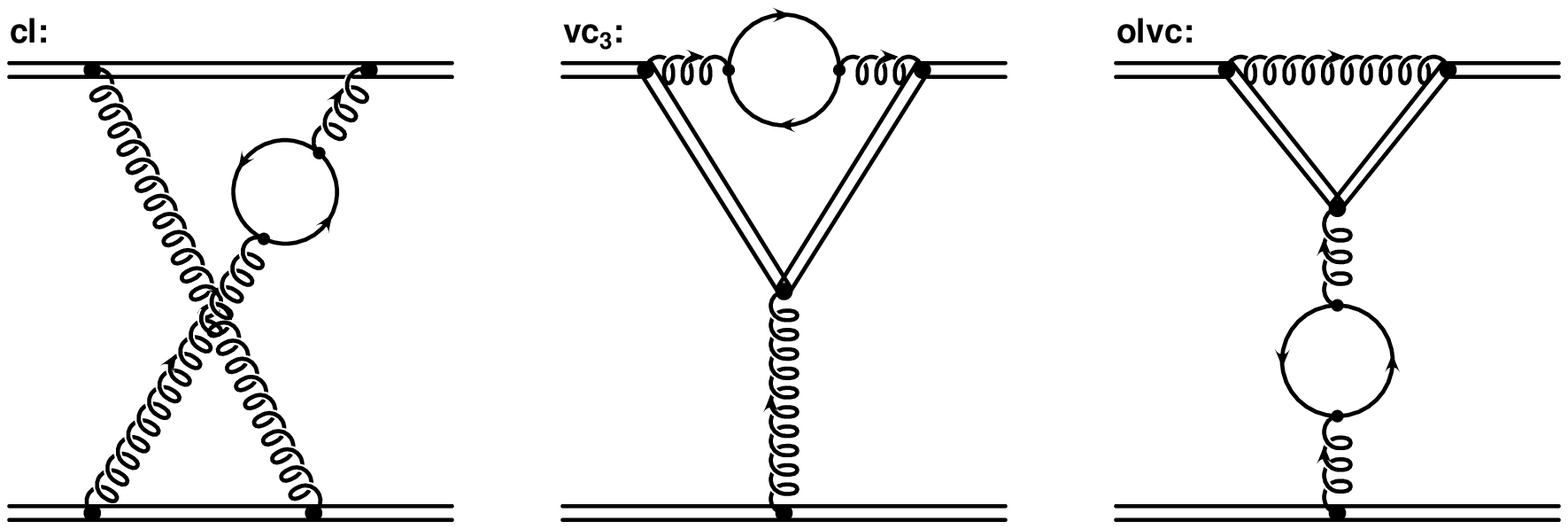,width=7.0cm}
\vspace{0.5cm} \\
\epsfig{file=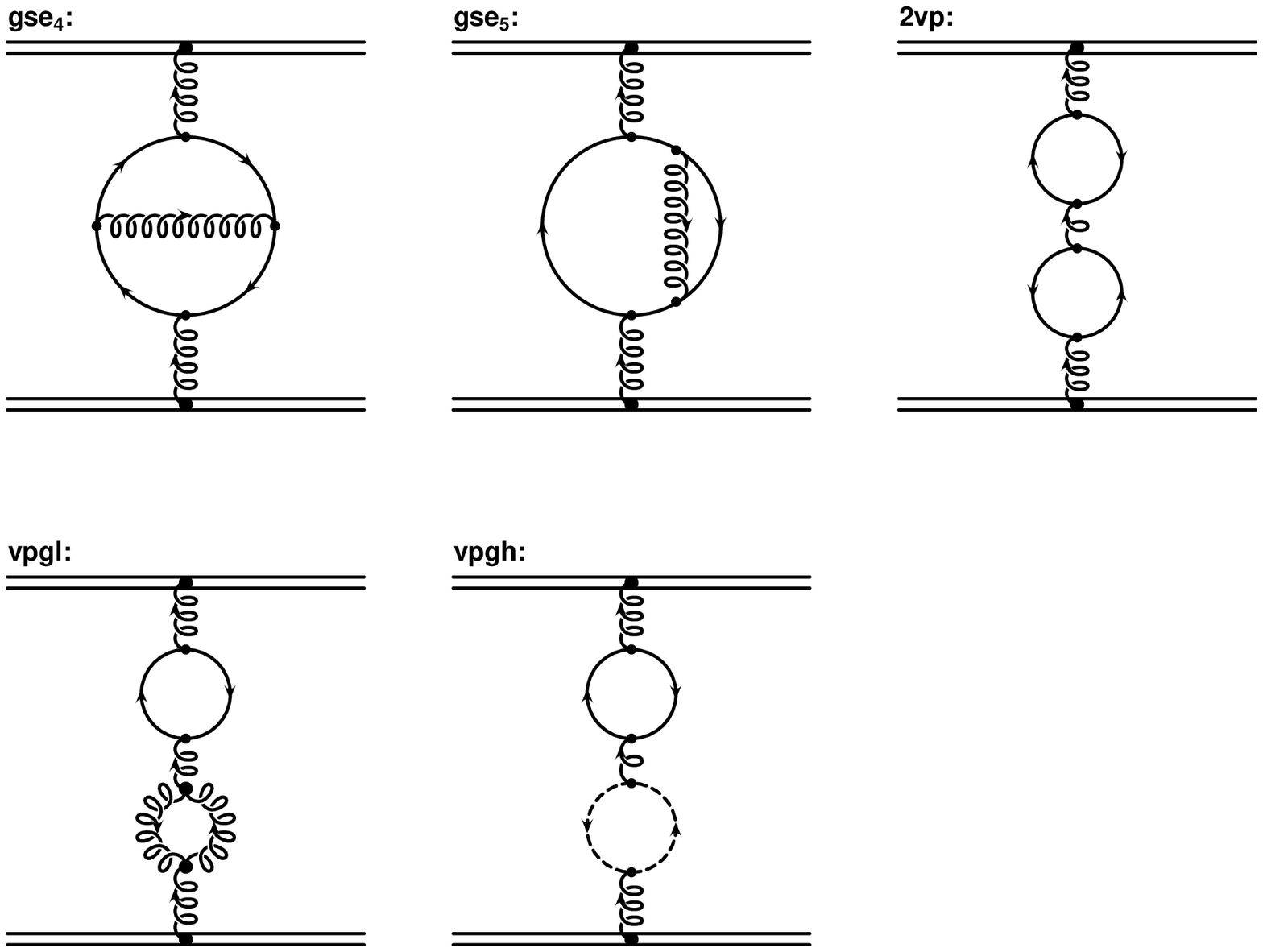,width=7.0cm}
\caption{The massive fermionic corrections to the heavy quark potential through
two loops in the Feynman gauge. The straight ladder diagram does not contribute
as it is already contained in the iteration of lower order amplitudes. The
middle line contains IR-divergent diagrams, however, their sum is IR-finite.
Contributions proportional to $C_F$ and $C_A$ are separately gauge invariant
for $\alpha_V$. After inclusion of the counterterms in Fig. \ref{fig:hqpct} the correct
massless limit given in
Refs. \cite{mp,ys} is obtained.
Details can be found in Ref. \cite{me1}.} \label{fig:tld}
\end{figure} 
\end{center}
\vspace{-0.8cm}
\section{Two loop corrections}

\begin{center}
\begin{figure}[tb]
\epsfig{file=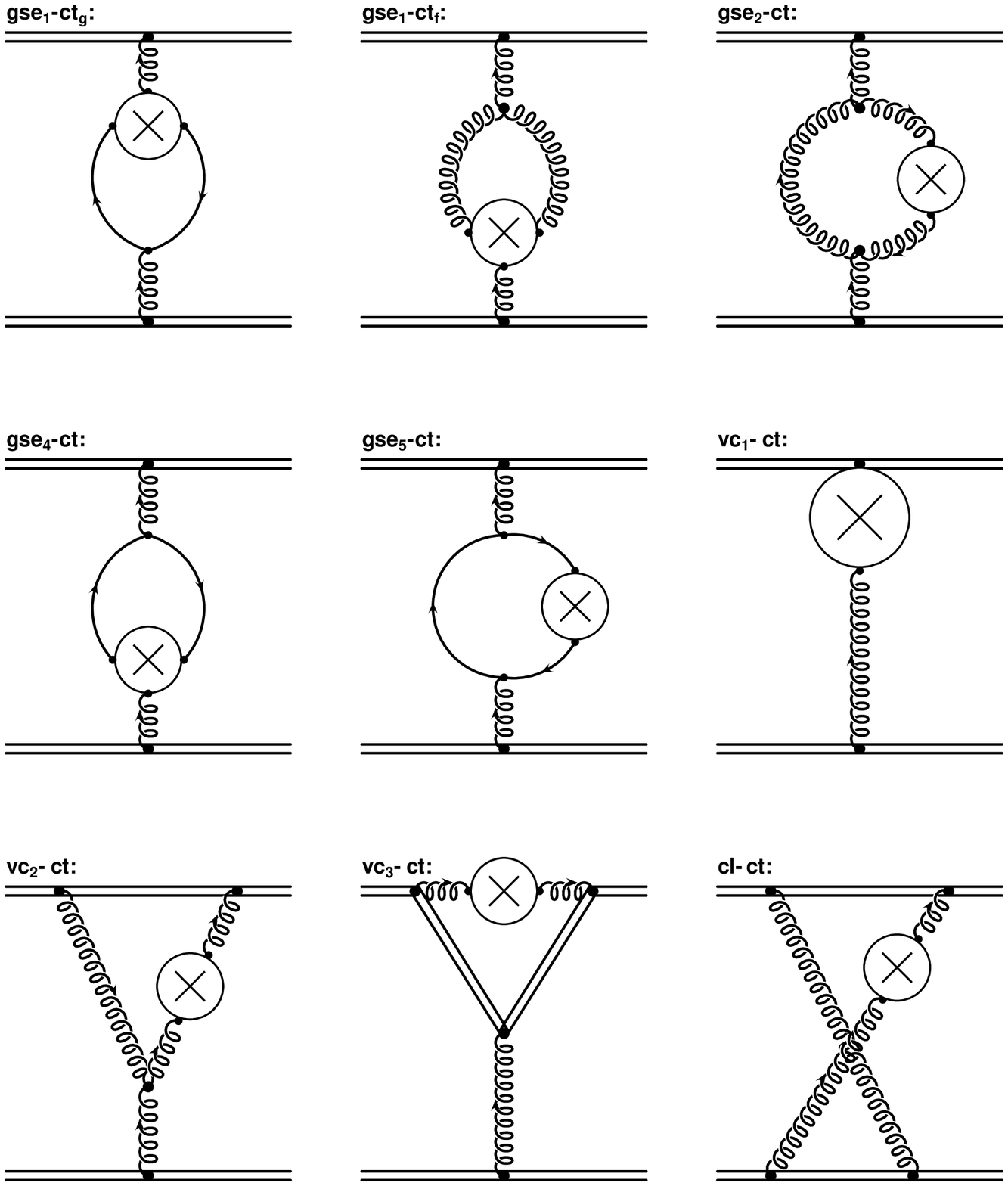,width=7.0cm}
\caption{The two loop counterterms corresponding to the diagrams in Fig.
\ref{fig:tld}.
Adding these contributions to the original graphs removes all non-local
functions from the occurring pole terms. The only exception are $\frac{m^2}{
\epsilon}$ terms in the two point functions which only cancel in the sum
of all two point diagrams. The fact that the tadpole
diagram has no counterterm is already indicative of this cancellation.}
\label{fig:hqpct}
\end{figure}
\end{center}

The results obtained in Ref. \cite{me1} express the physical charge $\alpha_V$ in the MS-scheme,
which is used as a calculational tool,
in the following way:
\begin{eqnarray}
\alpha_V(Q,m) \!\!\!&=&\!\!\! \alpha_{{\mbox{\tiny MS}}}(\mu)
\left( 1 + v_1 (Q,m(\mu),\mu) \frac{
\alpha_{{\mbox{\tiny MS}}}(\mu)}{\pi} 
\right. \nonumber \\ && \!\!\! \left. + v_2 (Q,m(\mu),\mu)
\frac{\alpha^2_{{\mbox{\tiny MS}}}(\mu)}{\pi^2} + \cdots \right)
\label{eq:aVmu}
\end{eqnarray}
where $v_2$ contains the diagrams of Fig. \ref{fig:tld} and the MS-counterterms displayed in Fig.
\ref{fig:hqpct}. A strong check of the results in Ref. \cite{me1} is given by the successful 
reproduction of the fermionic gluon wave function renormalization constant (RC) and the locality
of all other RC's as these are mass independent in minimally subtracted schemes.

For the heavier quark masses $m_c$, $m_b$ and $m_t$ the pole-mass definition
is suitable and allows for a straightforward Abelian limit as well as the renormalization scale
independence of the Gell-Mann Low function below.
The next-to-leading order
relation between the MS mass $m(\mu)$ and the pole mass $m$
is given by \cite{t}:
\begin{eqnarray}
m(\mu)&=&m \left[ 1 - C_F \frac{\alpha_{{\mbox{\tiny MS}}}(\mu)}{\pi}
\left( 1 + \frac{3}{2} \log \frac{
\mu}{m} - \right. \right. \nonumber \\ && \left. \left.
\frac{3}{4}\left[\gamma - \log ( 4 \pi)\right] \right) \right]
\label{eq:mrm}
\end{eqnarray}
where $\gamma$ is the Euler constant.
Inserting Eq.~(\ref{eq:mrm}) into  Eq.~(\ref{eq:aVmu}) gives at
next-to-next-to-leading order
\begin{eqnarray}
\alpha_V(Q,m) &=& \alpha_{{\mbox{\tiny MS}}}(\mu)
\left[ 1 +
v_1 (Q,m,\mu) \frac{\alpha_{{\mbox{\tiny MS}}}(\mu)}{\pi} +
\right. \nonumber \\ &&\!\!\!\!\!\!\!\!\!\!\!\!\!\!\!\!\!\!\!\! \left.
\left[v_2 (Q,m,\mu) + \Delta_m(Q,m,\mu)\right]
\frac{\alpha_{\mbox{\tiny MS}}^2(\mu)}{\pi^2}
\right] \label{eq:aVmupm}
\end{eqnarray}
where $\Delta_m(Q,m,\mu)$ denotes the contribution arising from $v_1$
when changing from the MS mass to the pole mass.

\section{Numerical solutions of the Gell-Mann Low function}

The Gell-Mann Low function \cite{gl} for the $V$-scheme is defined as the
total logarithmic
derivative of the effective charge with respect to the physical
momentum transfer scale $Q$:
\begin{equation}\label{eq:psiv}
\Psi_V \!\! \left( \! \frac{Q}{m} \! \right) \! \equiv \! \frac{d \alpha_V (Q,m)}{
d \log Q} \! \equiv \! \sum^{\infty}_{i=0} \! -\psi_{V}^{(i)} \frac{\alpha_V^{i+2} (Q,m)}
{\pi^{i+1}} 
\end{equation}
\begin{center}
\begin{figure}[tb]
\epsfig{file=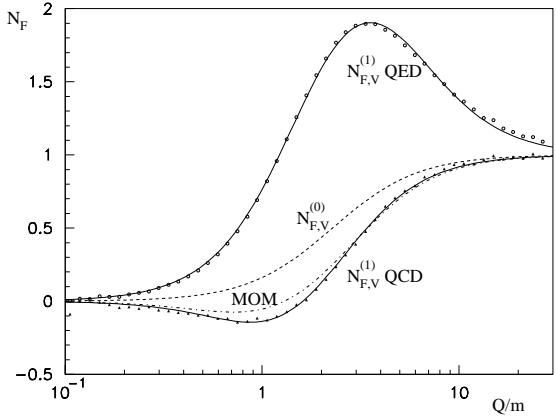,width=7.5cm}
\caption{The numerical results for the gauge-invariant  $N_{F,V}^{(1)}$ in QED
(open circles) and QCD (triangles) with the best $\chi^2$ fits of
Eqs.~(\ref{eq:psi1qed}) and (\ref{eq:psi1qcd})
superimposed respectively.   The dashed
line shows the one-loop $N_{F,V}^{(0)}$ function.
For comparison we
also show the gauge dependent two-loop result obtained in MOM schemes
(dash-dot)  \protect\cite{yh,jt}. At large $\frac{Q}{m}$ the theory becomes
effectively massless, and both schemes agree as expected. The figure also
illustrates the decoupling of heavy quarks at small $\frac{Q}{m}$.}
\label{fig:nfV}
\end{figure}
\end{center}
For the massive case all the mass effects will be collected into a
mass-dependent function $N_F$. In other words we will write
\begin{eqnarray}\label{eq:nfv0}
\psi_{V}^{(0)}\left( \frac{Q}{m}  \right) &=&
\frac{11}{2}-\frac{1}{3}N_{F,V}^{(0)}\left( \frac{Q}{m} \right) \\
\label{eq:nfv1}
\psi_{V}^{(1)} \left(  \frac{Q}{m}  \right)&=&
 \frac{51}{4} -\frac{19}{12}N_{F,V}^{(1)}\left( \frac{Q}{m} \right) 
 \end{eqnarray}
 where the subscript $V$ indicates the scheme dependence of
 $N_{F,V}^{(0)}$ and $N_{F,V}^{(1)}$.

 Taking the derivative of  Eq.~(\ref{eq:aVmupm}) with respect to $\log Q$ and
 re-expanding the result in  $\alpha_V(Q,m)$ gives the following equations
 for the first two coefficients of $\Psi_V$:
 \begin{eqnarray}
 \psi_{V}^{(0)}\left( \frac{Q}{m} \right) &=& -\frac{d v_1 (Q,m,\mu)}
 {d \log Q} \label{eq:psi0} \\
 \psi_{V}^{(1)} \left( \frac{Q}{m} \right)&=&
 -\frac{d [v_2 (Q,m,\mu)+ \Delta_m(Q,m,\mu)]}{d \log Q} \nonumber \\ && + 2
 v_1 (Q,m,\mu) \frac{d v_1 (Q,m,\mu)}{d \log Q} 
 \label{eq:psi1}
 \end{eqnarray}
 The argument $Q/m$  indicates that there is no
 renormalization-scale  dependence in Eqs.~(\ref{eq:psi0}) and
 (\ref{eq:psi1}). Rather, $\psi_{V}^{(0)}$ and $\psi_{V}^{(1)}$ are functions
 of the ratio of the physical momentum transfer
 $Q = \sqrt{-q^2}$ and the pole mass $m$ only.
A numerical solution based on the MC-integrator VEGAS and
numerical differentiation gives stable results summarized in Fig. \ref{fig:nfV}.
In the case of QCD we obtain the following approximate form for the
effective number of flavors for a given quark with mass $m$ \cite{me2}:
\begin{equation}
N^{(1)}_{F,V} \left( \frac{Q}{m} \right) \approx \frac{\left( -0.571 + 0.221
\displaystyle\frac{Q^2}{m^2} \right)
\displaystyle\frac{Q^2}{m^2}}{1+1.326 \displaystyle\frac{Q^2}{m^2} + 0.221
\displaystyle\frac{Q^4}{m^4}}
\label{eq:psi1qcd}
\end{equation}
and for QED
\begin{equation}
N^{(1)}_{F,V} \left( \frac{Q}{m} \right) \approx \frac{\left( 1.069 + 0.0133
\displaystyle\frac{Q^2}{m^2} \right)
\displaystyle\frac{Q^2}{m^2}}{1+0.402 \displaystyle\frac{Q^2}{m^2} + 0.0133
\displaystyle\frac{Q^4}{m^4}} 
\label{eq:psi1qed}
\end{equation}
The results of our numerical calculation of $N_{F,V}^{(1)}$ in the
$V$-scheme for QCD and QED are shown in Fig.~\ref{fig:nfV}.
The decoupling of heavy quarks becomes manifest at small $Q/m$, and
the massless limit is attained for large $Q/m$.
\begin{center}
\begin{figure}[p]
\epsfig{file=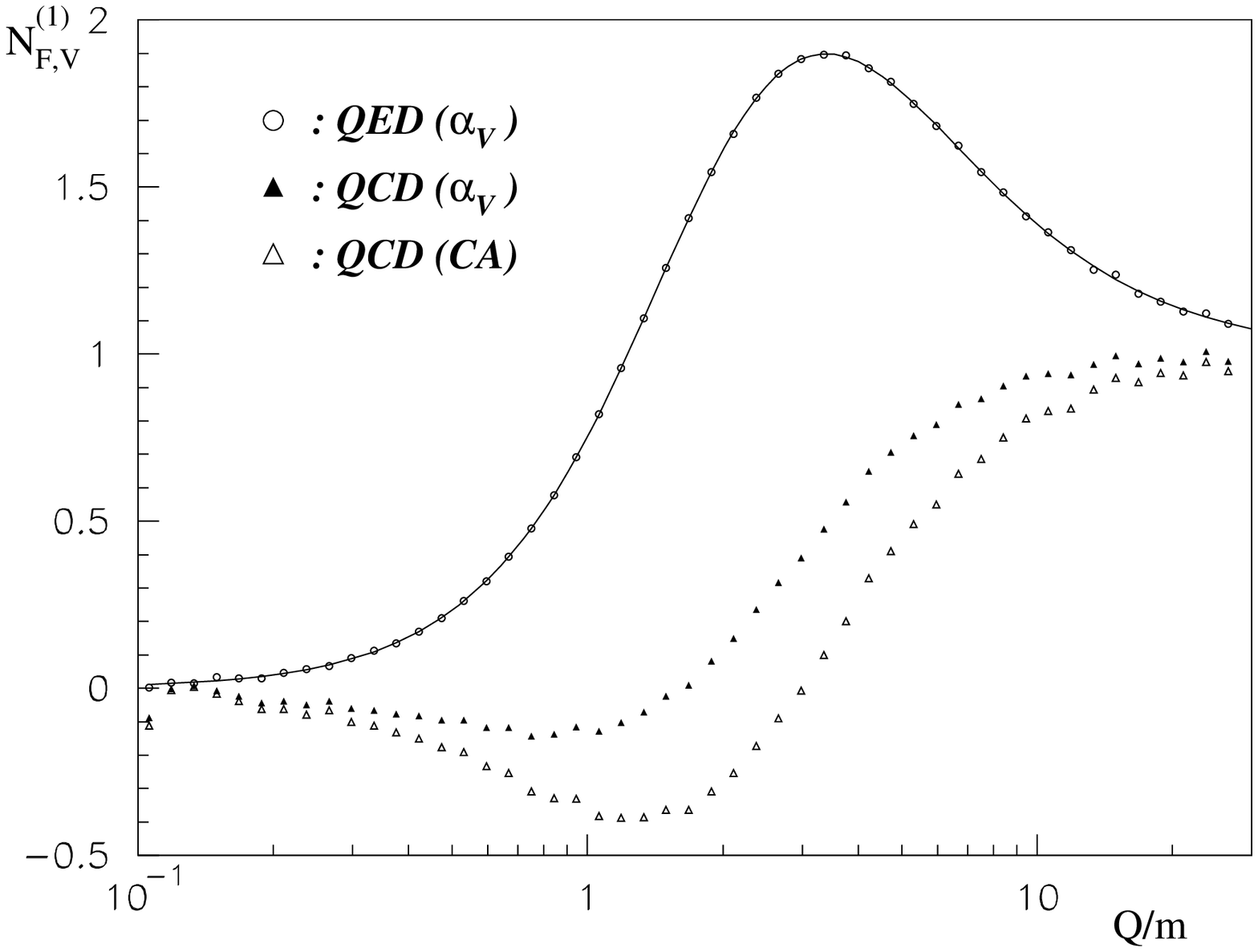,width=7.0cm}
\epsfig{file=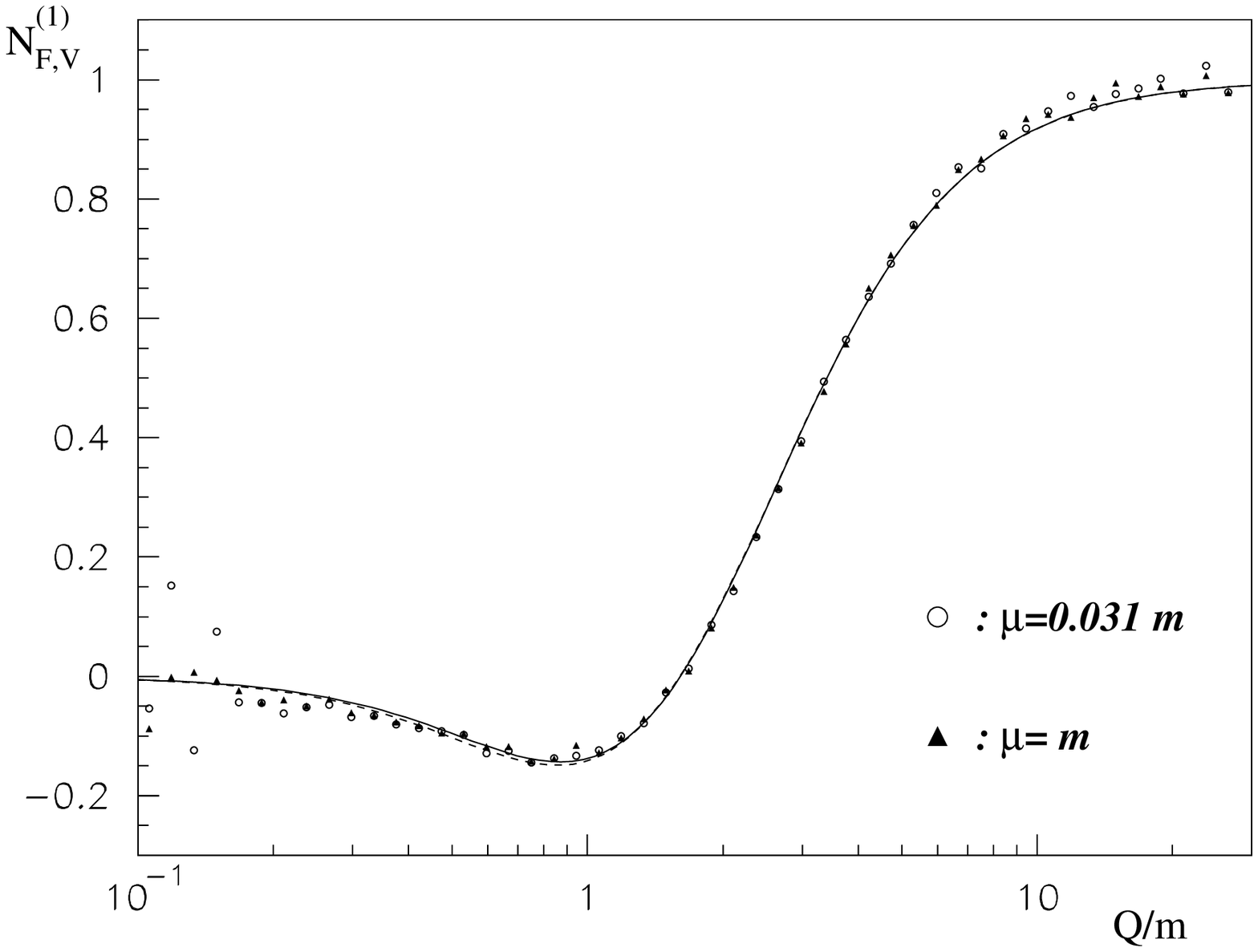,width=7.0cm} \vspace{-1cm} \\
\caption{The upper diagram displays a
comparison of the Abelian limit of our results (open circles)
for $N_{F,V}^{(1)}$  based on the calculation in
Ref.~\protect\cite{me1} which was done
in the MS-scheme with the well known result in
the literature \protect\cite{ks} done in the on-shell renormalization
scheme (solid line).  Also shown are the gauge invariant non-Abelian
contribution only ($\propto C_A$) (open triangles) as well as the sum of all
terms in QCD (solid triangles). The correct Abelian behavior is a very strong
check on the results given in Ref.~\protect\cite{me1}. The lower diagram
illustrates the renormalization scale independence of
the two-loop effective number of flavors $N_{F,V}^{(1)}$  as a function of the
ratio of the physical momentum transfer $Q$ over the pole mass $m$.  Numerical
instabilities  are visible for small values of $\frac{Q}{m}$ and occur because
of limited Monte Carlo statistics ($10^7$ evaluations for each
of the 50 iterations).  The two fits obtained, which agree within statistical
errors, are shown as a solid and dashed line for $\mu=m$ and $\mu=0.031m$
respectively.}
\label{fig:test}
\end{figure}
\end{center}
\begin{center}
\begin{figure}[p] $ $ \vspace{-1cm} \\
\epsfig{file=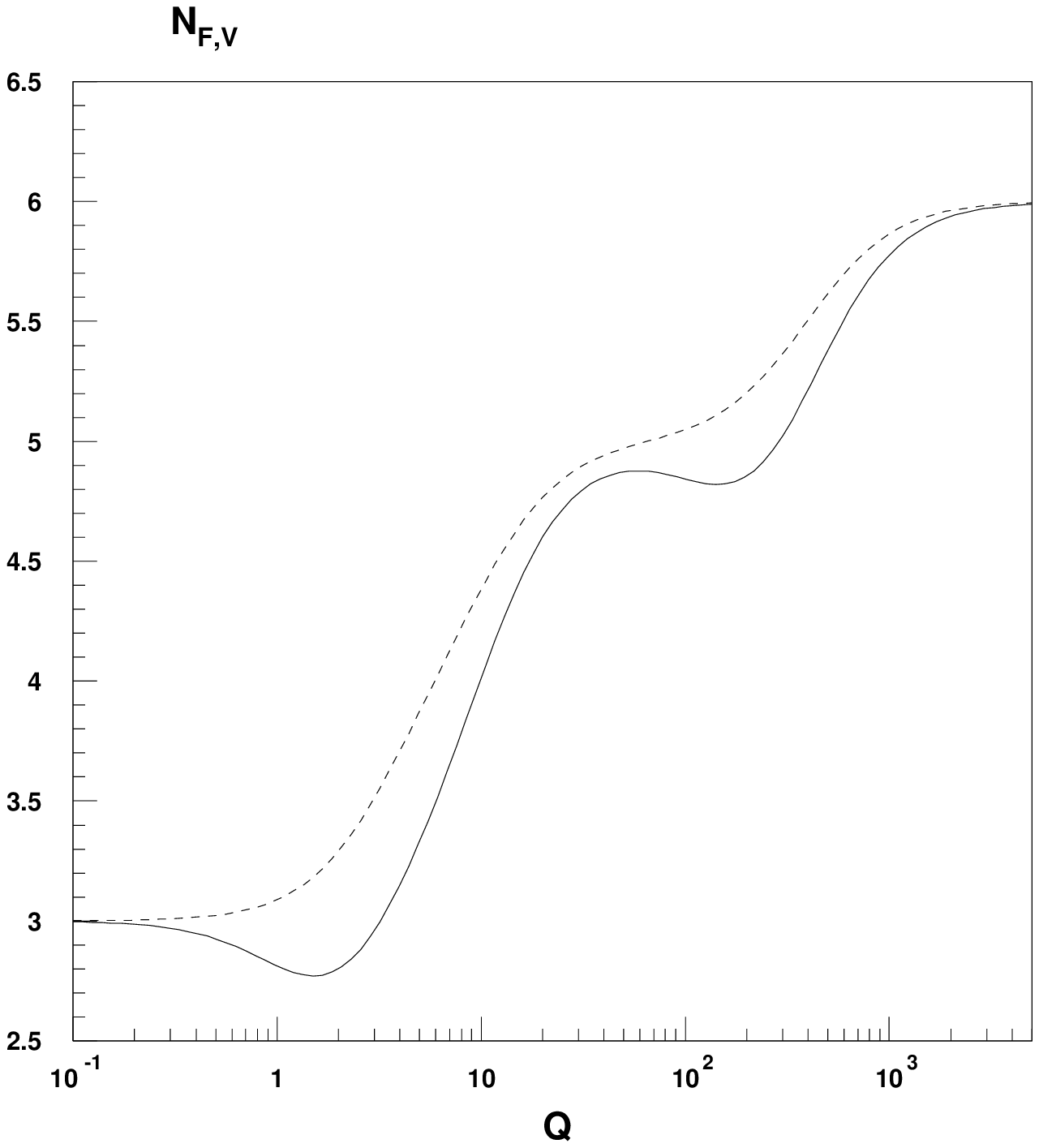,width=7.5cm} 
\epsfig{file=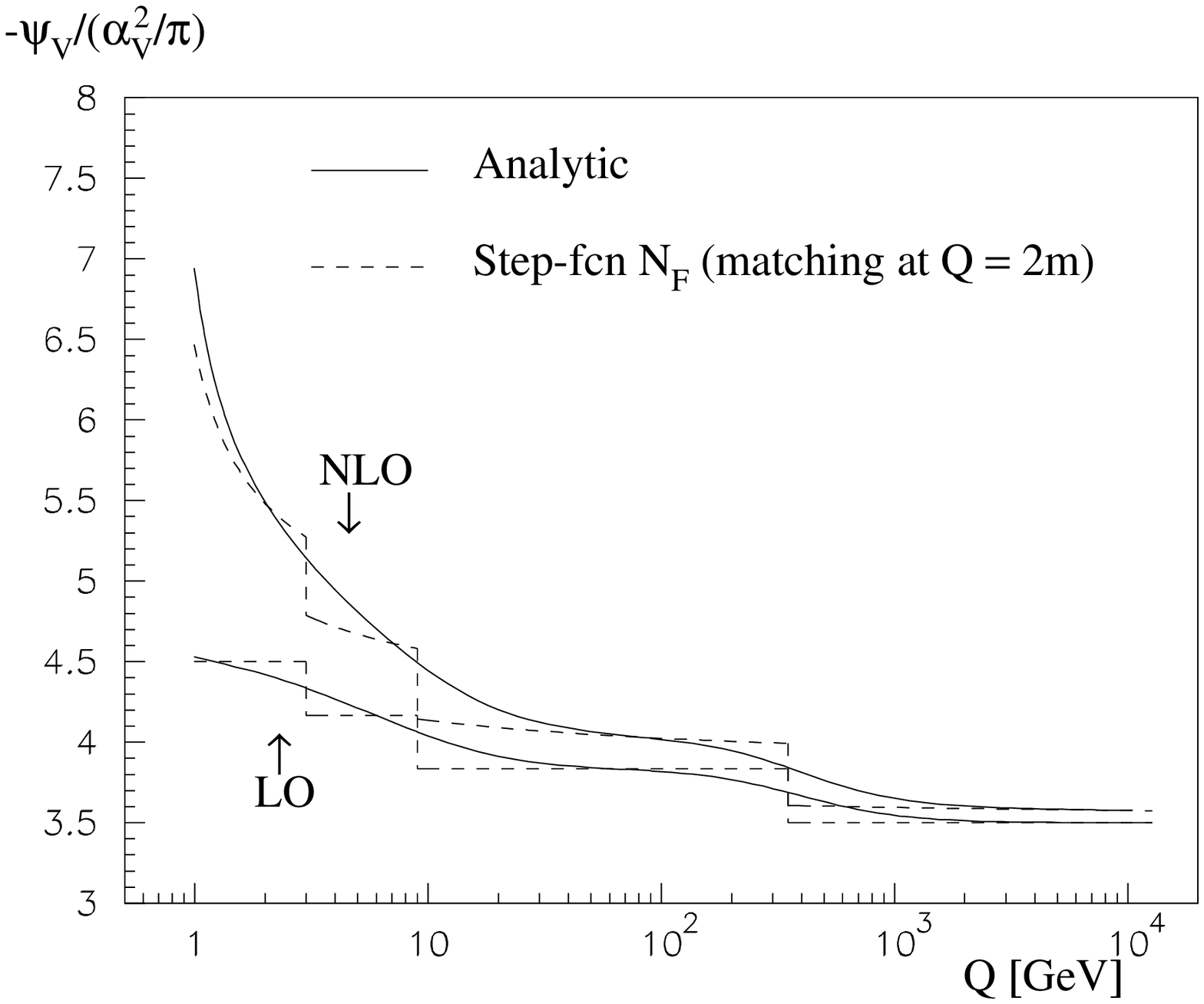,width=7.0cm} \vspace{-1cm} \\
\caption{The upper plot shows the sum of the effective number of flavors for one (dashed line) and
two loops in the V-scheme. We use quark pole masses with $m_c=1.5$GeV, $m_b=
4.5$GeV and $m_t=173$GeV. The two loop $N^{(1)}_{F,V}$ starts to decrease from the
fixed starting point $N_{F,V}=3$ due to
the novel non-Abelian anti screening corrections and then increases more 
rapidly as the one 
loop $N^{(0)}_{F,V}$. Below 1TeV, there is no regime for which the quark masses
can be neglected.
The lower plot displays the  scaled $\Psi$-function,
$- {\Psi}_{V}/({\alpha}^2_{V}/\pi)$ in the analytic V-scheme
${\alpha}_{V}(Q,m_i)$ (solid)
compared to the $\alpha_{V}(Q,\Theta)$  scheme with
discrete theta-function treatment of flavor
thresholds with continuous matching at $Q=m$ (dashed).}
\label{fig:nfsum}
\end{figure}
\end{center}
\vspace{-1.5cm}
We can also apply the same fitting procedure to the dependence of
the one-loop effective
$N^{(0)}_{F,V} \approx \frac{1}{1+5.19 \frac{m^2}{Q^2}}$.
Fig. \ref{fig:test} demonstrates that the new non-Abelian contributions ($\sim C_A$) are
responsible for the negative $N^{(1)}_{F,V}$ at intermediate $Q/m$ due to anti-screening.
The Abelian corrections on the other hand are larger than 1 in this regime and agree
with the literature results obtained in the on-shell scheme \cite{ks}.
In addition the lower graph of Fig. \ref{fig:test} demonstrates the renormalization scale
independence of the solution to the Gell-Mann Low function.

Fig. \ref{fig:nfsum} demonstrates the smoothness and analyticity of the renormalization group
solutions and compares the massive results with the massless ones including one-loop matching
at the two loop order. The figure demonstrates that there is really no regime below 1 TeV
where quarks can be considered massless for running coupling effects in the V-scheme.

\section{Conclusions}

In summary, we have presented the gauge invariant mass dependence of $\alpha_s$ through two
loops in the physically motivated V-scheme. The result was shown to posses the correct massless
limit and gives automatic decoupling of heavy quarks. In addition the correct Abelian limit
is reproduced and the renormalization scale independent results can be parameterized by a simple
analytical fitting function. Non-Abelian anti-screening effects lead to a negative number of
flavors for intermediate energies at the two loop level.

Massive renormalization group solutions are
scheme dependent already at one-loop, however, the mass dependence of $\alpha_V$ can be
transferred to other physical charges through commensurate scale relations \cite{bl}. In
Ref. \cite{me2} this was done for the non-singlet hadronic width of the Z-boson and compared
with the $\overline{MS}$-scheme higher twist corrections. For perturbative energies a
persistent $\sim$1\% deviation was observed which characterizes the residual scheme dependence
of the two loop predictions for this observable. In a similar way, all other effective
charges can be described.

Other possible applications include the effect of a massive charm on the bottom mass
determination. Here massive charm corrections to the potential and in the running coupling
could potentially lead to a shift in the bottom mass of ${\cal O} \left( m_c \alpha_s^2(m_b) \right)$
and thus would need to be included into a proper analysis.
Also top quark physics at the NLC could provide fruitful ground for a V-scheme analysis. 

A further interesting comparison could be performed with lattice analyses investigating
the transition region of perturbative and non-perturbative regimes. For this purpose
the Fourier transform $\alpha_V(1/r)=\int \frac{d^3q}{(2\pi)^3} \alpha_V (q^2,m^2)$ or
$\alpha_{q,\overline{q}}(r) \equiv -r^2 \frac{\partial V(1/r)}{\partial r}$ must be
obtained.

\section*{Acknowledgments}

I would like to thank my collaborators S.J.~Brodsky and J.~Rathsman for their
contributions to the results presented here.

\end{document}